\def\year{2022}\relax
\documentclass[letterpaper]{article} %
\usepackage{aaai22}  %
\usepackage{times}  %
\usepackage{helvet}  %
\usepackage{courier}  %
\usepackage[hyphens]{url}  %
\usepackage{graphicx} %
\usepackage{natbib}  %
\usepackage{caption} %
\DeclareCaptionStyle{ruled}{labelfont=normalfont,labelsep=colon,strut=off} %
\usepackage{algorithm}
\usepackage{algorithmic}

\usepackage{newfloat}
\usepackage{listings}
\floatstyle{ruled}
\newfloat{listing}{tb}{lst}{}
\floatname{listing}{Listing}
\nocopyright
\usepackage{graphicx}
\usepackage{amsmath}
\usepackage{bm}
\newcommand{\bs}{\boldsymbol}
\usepackage{amssymb}
\usepackage{multirow}
\usepackage{booktabs}
\usepackage{enumitem}

\title{Semi-Structured Query Grounding for Document-Oriented Databases\\ with Deep Retrieval and Its Application to Receipt and POI Matching}
\author {
    Geewook Kim\textsuperscript{\rm a,}\thanks{Corresponding author: gwkim.rsrch@gmail.com},
    Wonseok Hwang\textsuperscript{\rm b,}\thanks{This work was done while the authors were at NAVER Corp.},
    Minjoon Seo\textsuperscript{\rm c,\dag},
    Seunghyun Park\textsuperscript{\rm a}
}
\affiliations {
    \textsuperscript{\rm a} Clova AI Research, NAVER Corp.\\
    \textsuperscript{\rm b} LBox Co., Ltd.\\
    \textsuperscript{\rm c} Korea Advanced Institute of Science and Technology\\
}

\begin{document}

\maketitle

\begin{abstract}
Semi-structured query systems for document-oriented databases have many real applications.
One particular application that we are interested in is matching each financial receipt image with its corresponding place of interest (POI, e.g., restaurant) in the nationwide database.
The problem is especially challenging in the real production environment where many similar or incomplete entries exist in the database and queries are noisy (e.g., errors in optical character recognition).
In this work, we aim to address practical challenges when using embedding-based retrieval for the query grounding problem in semi-structured data.
Leveraging recent advancements in deep language encoding for retrieval, we conduct extensive experiments to find the most effective combination of modules for the embedding and retrieval of both query and database entries without any manually engineered component.
The proposed model significantly outperforms the conventional manual pattern-based model while requiring much less development and maintenance cost.
We also discuss some core observations in our experiments, which could be helpful for practitioners working on a similar problem in other domains.
\end{abstract}

\begin{figure}[t]
    \centering
    \includegraphics[width=\linewidth]{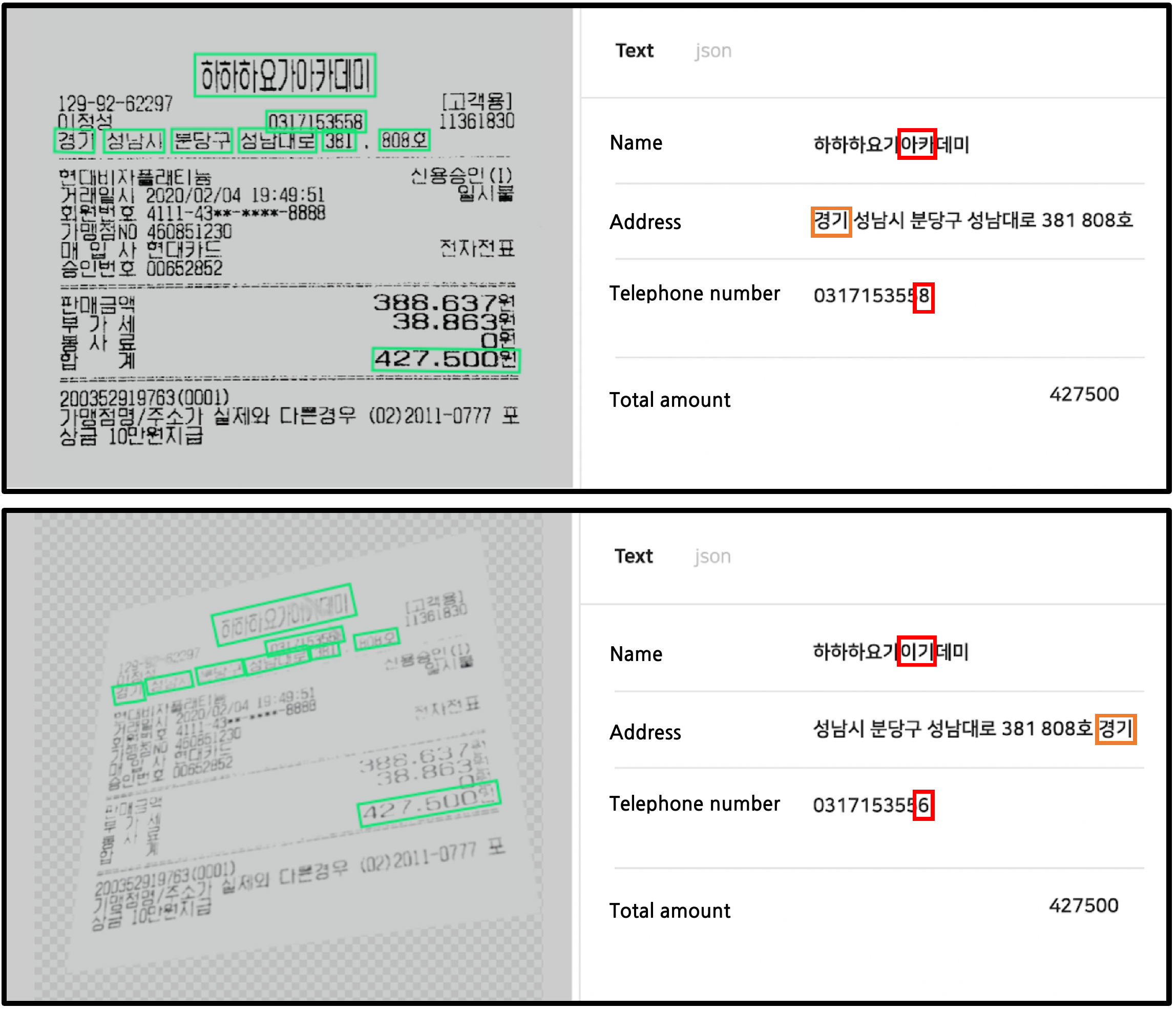}
    \caption{\textbf{Results of our OCR and Parsing web demo\footnotemark.} The system successfully extracts structured information of the input paper bill image (upper). However, given a noisy sample (we added common noise to the sample, e.g., low image resolution, perspective view and blur), text recognition (red) and word order prediction (orange) failed (bottom).} %
    \label{fig:ocr_demo}
\end{figure}

\begin{figure*}[t]
    \centering
    \includegraphics[width=\linewidth]{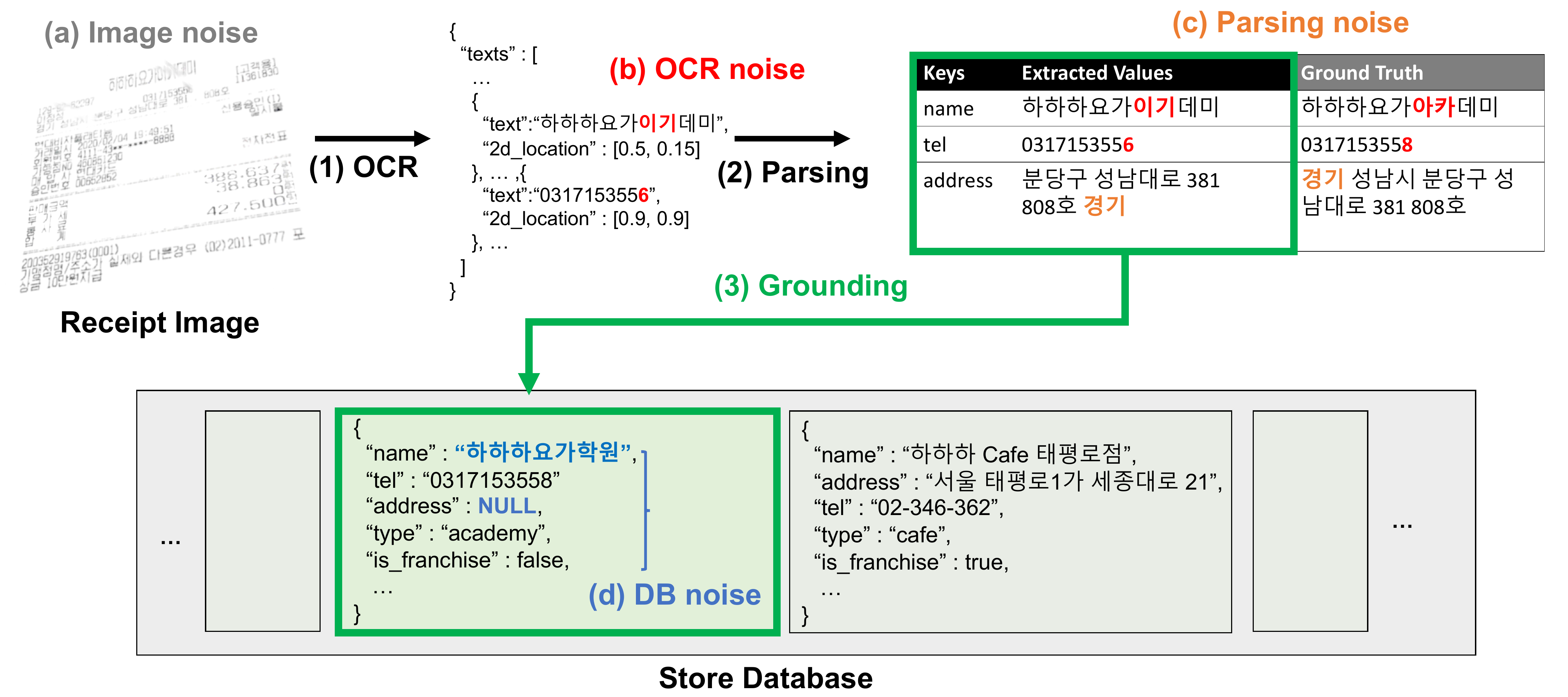}
    \caption{\textbf{An overview of \emph{Place of Interest (POI) Match}}. Given an input receipt image, (1) an OCR engine is applied to extract texts and their 2D locations, (2) the extracted information is parsed to a semi-structured representation (e.g., in JSON format), and (3) the input receipt is grounded to a POI database entry. Grounding is a difficult task since (a) the quality of the input image can be poor, (b) OCR noise or (c) Parsing noise (e.g., the order of words in the address is incorrectly estimated in the example) can lead to wrong information, and finally (d) the extracted information may not exactly match the information in the database (e.g., missing values or outdated information).}
    \label{fig:teaser}
\end{figure*}

\section{Introduction}

Querying a database with semi-structured natural language (e.g., in JSON format) has many real-world applications~\citep{arora2013modeling, mongodbcouchdb}.
One particular application that we are interested in is \emph{\textbf{Place of Interest (POI) Match}}, which is an active product where the user input is a paper bill image and the desired output is a database entry that corresponds to the POI where the financial transaction is deemed to have taken place.
Most commercial receipts have place information, and thus the receipt and POI matching can be done by extracting the information from the receipt image and querying the POI database.
This can be done in steps as follows:
(1) An optical character recognition (OCR) is first applied to extract text and its geometric location from the receipt image.
(2) A document parsing (information extraction) process is applied to determine which pre-defined category (e.g., store name and store address) the obtained text belongs to (see the details in Figure~\ref{fig:ocr_demo}).
(3) Finally, the grounding process uses the extracted information as a query to the POI database, i.e., find the corresponding store in the database. We present the whole end-to-end pipeline for \emph{POI Match} in Figure~\ref{fig:teaser}.
This paper focuses on the \emph{grounding} stage of the pipeline, or more generally, the problem of grounding a semi-structured query on document-oriented databases, and we use \emph{POI Match} as the primary testbed for evaluating our solution to the problem.

\footnotetext{\url{https://clova.ai/ocr}}
One baseline approach for the grounding module (that has been in service in our product for the last few years) is to build a rule-based system with document-oriented databases~\citep{mongodb,couchdb}.
In the system,
querying functions implemented in the databases are used to find the store, e.g., regular expression-based search algorithms~\citep{10.1145/1376616.1376706, lockard-etal-2019-openceres, nadig-etal-2020-database}.
If multiple stores are retrieved by the function, 
the system can finalize the answer with predefined rules,
e.g., return the most visited store.
Although the idea is simple, it is difficult to make good rules to cover numerous user queries.
However, in fact, such a heuristic rule-based approach is fundamentally limited as there are limitless exceptions in real-world environments.

In practice, the problem becomes increasingly non-trivial when (1) the input image is of bad quality, so that the model fails to extract correct information (See the bottom of  Figure~\ref{fig:ocr_demo} and (a, b, c) of Figure~\ref{fig:teaser}), or (2) the information in the receipt does not exactly match the corresponding database entry, e.g., missing value or outdated information (See (d) of Figure~\ref{fig:teaser}).
Furthermore, the size of a database is often in the order of millions or more (for instance, our database for \emph{POI Match} has approximately six million entries) and many database entries include similar values, such as the name of franchises, so that it is difficult to differentiate.
Hence, for robust grounding in a real-world scenario, the model should consider multiple fields simultaneously and should handle both semantic and lexical similarities among the data to distinguish similar entries.
Also, the model should deal with the issues of missing values and noises of the data.
Because of these difficulties, our current rule-based model has a more than 30\% failure rate for all incoming queries.

To address these issues, we combine an idea of embedding-based retrieval~\citep{karpukhin-etal-2020-dense,fairembed} into database querying.
In our system, both queries and database entries are represented by vectors.
Given a query, the corresponding database entries are retrieved by searching nearest neighbors in the vector space.
To make a good system for grounding,
we first introduce a framework for grounding that consists of several replaceable modules.
And then, we conduct extensive experiments to assess the contribution of individual modules more rigorously and find the best module combination over the model architectures.
The proposed model successfully alleviates the issues of missing values and noises of the data, and significantly outperforms the rule-based baseline model by more than 8\% point of top-1 matching accuracy in \emph{POI Match} (this corresponds to approximately 2 million user queries per month).
While our work primarily focuses on one particular application (\emph{POI Match}), we believe that our findings can be easily extended to other real applications that rely on semi-structured queries on databases.

\section{Background}

\subsection{OCR and Parsing}

Information extraction (IE) on semi-structured document images is a core step towards automated document processing. In general, document IE systems consist of two stages: OCR and Parsing.

\subsubsection{OCR}
The OCR process consists of two sub-steps: detection and recognition. First, in the detection procedure, all text regions in the given image are predicted~\citep{baek2019craft}. The recognition module receives the detected image patch as its input and extracts all texts in the patch~\citep{baek2019STRcomparisons}.

\subsubsection{Post-OCR Parsing}
The parsing task can be interpreted as a named entity recognition (NER) task that maps texts to predefined categories based on the recognized text and location information. The task aims to extract the information in a structured form (See Figure 1)~\citep{hwang2019postocr,hwang-etal-2021-spatial}. However, the inference cannot always be accurate, especially in a real application environment. There have been various additional studies on how to correct the recognized information, which is still often inefficient and error-prone in practice~\citep{post_ocr, hwang-etal-2021-cost}.

\subsection{Database Querying} In document-oriented databases, the data is stored in the form of semi-structured data that consists of several key-value pairs, e.g., JSON or XML.
Database querying is done by retrieving corresponding database entries that satisfy the input conditions.
For instance, in MongoDB~\citep{mongodb}, the database entries whose the value of ``color'' is ``gold'' is retrieved by using the input query \texttt{find({color:"gold"})}.
Querying with multiple conditions is also supported, e.g., \texttt{find({color:"gold",type:"necklace"})}.

Database querying often fails in real-world applications since both queries and database entries are noisy in most practical scenarios, e.g., misspellings in a query or missing values in database entries.
A common solution to this problem is using a text-based search engine (e.g., Apache Lucene)~\citep{elasticsearch, couchdb}.
However, the performances are limited as the core idea is based on simple string distance calculation algorithms which are not good at capturing semantic similarities among the data~\citep{robertson2009probabilistic, elasticsearch}.

\subsection{Embedding-based Retrieval}
To retrieve textual data such as words, sentences, or documents,
many modern NLP applications convert the data into vector representations, i.e.,
embeddings~\citep{mikolov-etal-2013-linguistic,devlin-etal-2019-bert,karpukhin-etal-2020-dense}.
Once the data is embedded in vector space, 
retrieval can be done efficiently by calculating similarities among the vectors.
For instance, if the data is embedded into an inner-product space, the retrieval can be done with maximum inner product search algorithms~\citep{pmlr-v89-ding19a,tan-etal-2019-efficient, faiss}.
The embeddings are expected to hold characteristics, properties, or even semantics of the data so that retrieval targets can be found simply by calculating the distances in the vector space.
To obtain such good embeddings,
a range of representation learning methods has been studied~\citep{mikolov-etal-2013-linguistic,kim-2014-convolutional,bojanowski-etal-2017-enriching}
and most of the modern methods utilize BERT-based models to embed the data~\citep{devlin-etal-2019-bert,karpukhin-etal-2020-dense}.

\begin{figure}[t]
    \centering
    \includegraphics[width=\linewidth]{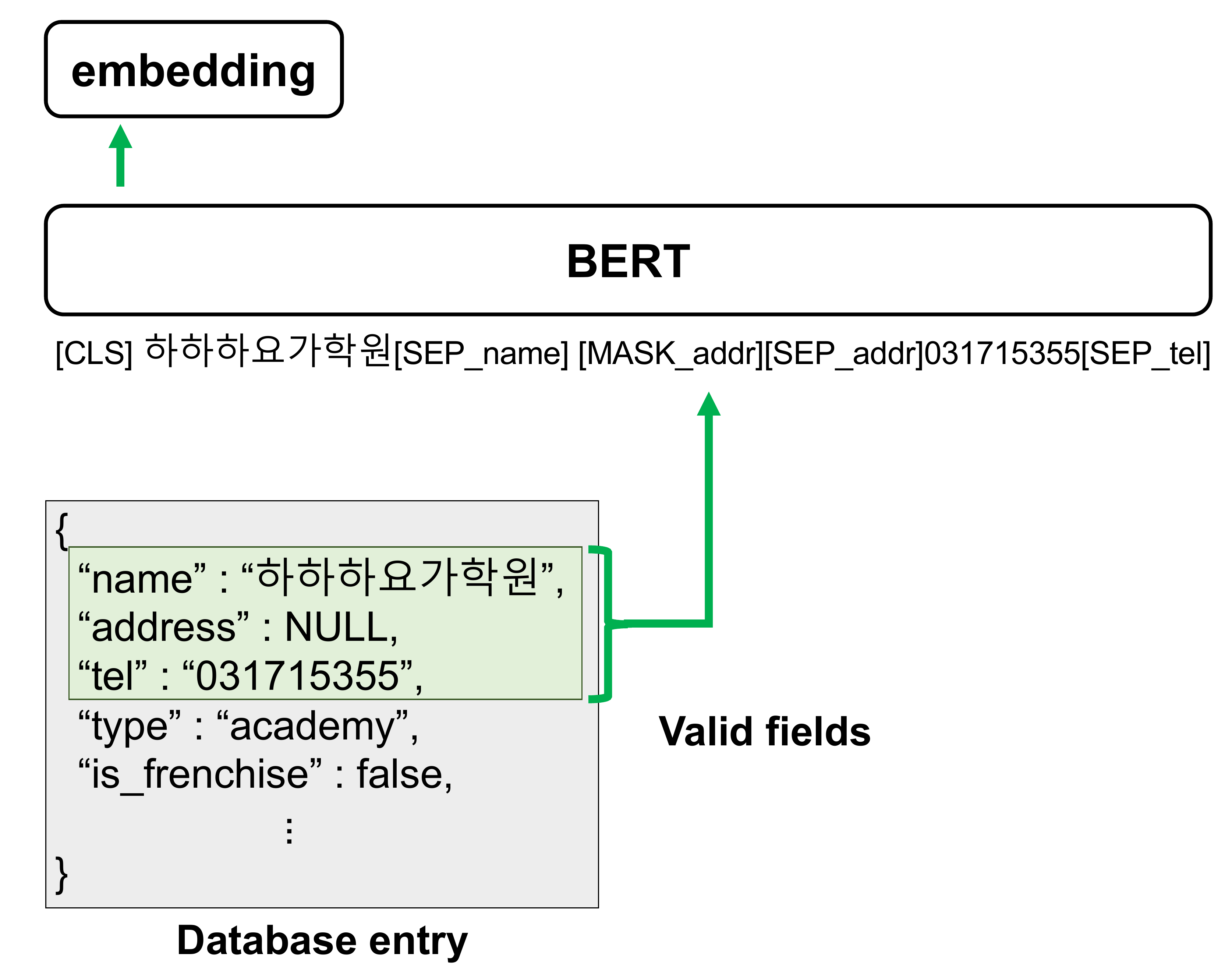}
    \caption{\textbf{An overview of semi-structured data embedding process.} Values of valid fields are concatenated and passed to a BERT-based model to obtain the vector representation.} %
    \label{fig:embedding_process}
\end{figure}

\section{Grounding}

In this section,
we aim to introduce our model for grounding.
Before introducing the model,
we formally define \emph{grounding} and provide some preliminaries.

\paragraph{Problem Definition.}
In this paper, we are interested in the task of finding a corresponding database entry from a large-scale document-oriented database, given a \textbf{noisy} semi-structured query (e.g., JSON). %
We denote this task as \emph{grounding}.
In the task, both queries and database entries are semi-structured data that consists of several key-value pairs.
We are given $n$ queries ($\in\mathcal{Q}$), $m$ database entries ($\in\mathcal{D}$),
and their associations which can be expressed as a matrix $(a_{i\hat{j}}) \in \mathbb{R}^{n\times m}$, where the value of $a_{i\hat{j}}$ represents the strength of the association between $i$-th query and $\hat{j}$-th database entry.
For instance, we may use the number of link (edge) between $i$-th query and $\hat{j}$-th database entry in the trainset as the association strength $a_{i\hat{j}}$.
The model is trained to predict the given association strengths $(a_{i\hat{j}})$.
In a test phase, for an input user query, the database entry with high predicted association strength is considered as the matching target.
We assume the target is a single database entry but it can be a set of database entries if the database is noised with duplicated entries (this is not rare in real application environments).

\subsection{Grounding Model}

We bring an idea of embedding-based retrieval into database querying.
In the proposed model, both queries and database entries are represented by dense vectors (i.e., embeddings).
Under this setting, grounding (database querying) becomes finding a close corresponding database entry vector for a given query vector.

The entire model architecture is based on a simple Siamese neural network~\citep{siamesenn, reimers-gurevych-2019-sentence,gillick-etal-2019-learning,wips,karpukhin-etal-2020-dense}.
Two neural networks (i.e., backbones) learn nonlinear mapping functions to learn the embeddings of queries and database entries that are used for database querying.
The querying is done by calculating similarity scores among the obtained input query vector and the embeddings of database entries.

\subsubsection{Backbones}
We use BERT-based models~\citep{devlin-etal-2019-bert,conneau-etal-2020-unsupervised} following recent trends in NLP. %
To apply BERT-based models on document-oriented data,
the first step is to vectorize the input data as the model assumes continuous data as inputs.
To prepare input vectors, we convert the document-oriented data into a sequence of tokens (See Figure~\ref{fig:embedding_process}).
Each token is converted to a vector representation through a look-up operation over a learnable embedding matrix~\citep{devlin-etal-2019-bert}.

\subsubsection{Input feature selection}
There are several choices to be made to make the input sequence;
(1) what fields to consider,
(2) how to concatenate the multiple field information. %
In our pipeline, a set of valid fields is first defined and only the values of the valid fields are concatenated to make a sequence (See Figure~\ref{fig:embedding_process}). 
In BERT-based models, a separator token \texttt{[SEP]} is used for the concatenation of different fields in general~\citep{devlin-etal-2019-bert,cohan-etal-2019-pretrained,karpukhin-etal-2020-dense}.
However, in the real-world data,
the number of fields is often more than two and the data has lots of missing values (i.e., null values).
In our preliminary experiments, we observed that a naive concatenation of multiple fields degrades the performance of the system.
To resolve this issue,
we introduce two additional special tokens (\texttt{[MASK]}$_{\ast}$ and \texttt{[SEP]}$_{\ast}$) per each field $\ast$.
The token \texttt{[MASK]}$_{\ast}$ is used when the value of the field $\ast$ is missing.
The token \texttt{[SEP]}$_{\ast}$ is used for separating the value of $\ast$ with other values.
See Figure~\ref{fig:embedding_process} for more details.

\subsubsection{Similarity functions}
To calculate the association strength $a_{i, \hat{j}}$ between queries and database entries,
the model learns $\boldsymbol{f}_{\boldsymbol\theta}:\mathcal{Q}\mapsto\mathcal{Y}$ and $\boldsymbol{g}_{\boldsymbol\vartheta}:\mathcal{D}\mapsto\mathcal{Y}$, where $\mathcal{Y} \in \mathbb{R}^{K}$ is a set where embeddings take a value with some dimensionality $K\in\mathbb{N}$.
The generated embeddings of query $\boldsymbol{y}_{i} := \boldsymbol{f}_{\boldsymbol\theta}(\bs q_i)$ and database entry $\boldsymbol{y}_{\hat{j}} := \boldsymbol{g}_{\boldsymbol\vartheta}(\bs d_{\hat{j}})$ capture the association strength by $s(\bs y_{i}, \bs y_{\hat{j}}) \propto a_{i, \hat{j}}$, where $s:\mathcal{Y}\times\mathcal{Y}\mapsto\mathbb{R}$ is a similarity function, such as, inner product similarity (IPS) $\langle\bs y_{i}, \bs y_{\hat{j}}\rangle$ or negative squared distance (NSD) $-||\bs y_{i} - \bs y_{\hat{j}}||^2_2$.

\subsection{Model Training}
Given the training data $\{a_{i,\hat j}\}$, 
the model parameters $\{\bs\theta, \bs\vartheta\}$ are learned by maximizing $\sum_{1\le i\le n, 1\le \hat{j}\le m} a_{i, \hat{j}} \log P(\hat{j}|i)$,
where $P(\hat{j}|i)$ is modeled as,
\begin{equation}
\frac{\exp(s(\bs y_{i}, \bs y_{\hat{j}}))}{\sum_{1\le \hat{k}\le m}\exp(s(\bs y_{i}, \bs y_{\hat{k}}))}.
\end{equation}
The summation in the denominator makes training difficult, especially when $m$ is large, which is often the case in real-world environments. 
The problem can be circumvented by using \textit{negative sampling},
leading to a modification of the above objective as follows,
\begin{equation}
\frac{\exp(s(\bs y_{i}, \bs y_{\hat{j}}))}{\sum_{\hat{k}\sim P_\text{neg}}\exp(s(\bs y_{i}, \bs y_{\hat{k}}))},
\end{equation}
where $P_{\text{neg}}$ is a distribution for negative sampling, such as, uniform, empirical frequency, etc.
For efficient model training,
we train the models with mini-batch gradient descent and use other samples in a mini-batch as negative samples~\cite{gillick-etal-2019-learning,karpukhin-etal-2020-dense}.

\subsection{Implementation}

We implement a library \texttt{grounder} that includes all fundamental functionality for training and deploying a grounding system.
\texttt{grounder} is implemented in PyTorch~\citep{pytorch} and built upon two public projects.
To use a range of BERT-based models, we use \texttt{transformers}~\citep{wolf-etal-2020-transformers} developed by huggingface.
For an efficient retrieval of nearest neighbors,
\texttt{faiss}~\citep{faiss} developed by Facebook AI is used. %
Our implementation will be publicly available on GitHub\footnote{\url{https://github.com/clovaai}}.

\section{Experiments}

In experiments, we study two architectures of backbone ({MBERT} and {XLMR}), two similarity functions ({IPS} and {NSD}),
two options in a separator token ({Single} and {Multi}), and
three options in masking missing values ({None}, {Single} and {Multi}).
All possible grounding module combinations (2×2×2×3= 24 in total) are evaluated to find the best module combination for our application \emph{POI Match}.
We also assess the efficacy of each module, and we believe that our findings in the experiments can be easily extended to other real-world applications that rely on grounding.

\paragraph{Common Settings.}

Given a receipt image and a database of stores,
our task is to find a corresponding store in the database that matches the image.
We evaluate all models with top-1 matching accuracy.
In experiments, we use 1 million receipt images and a database that contains approximately 6 million POI information.
Each query has 4 valid fields, where each field corresponds to the name, telephone number, address, and business number of the store respectively.
Each database entry has 4 valid fields, where each field corresponds to the name, telephone number, address, and street name (i.e., another type of address) of the store respectively.
There are many missing values in the data, for example, 21\% of the database entries have null values on the telephone number field and 17\% on the street name field respectively.
Each receipt image is linked to a database entry with the help of our rule-based model that has been in service in our application; the model is first applied to the image to find a corresponding database entry, and the links are refined by a human annotator.
The application is deployed in South Korea and the main language of the data is Korean.

\subsection{Comparison Models}

\subsubsection{Baseline}

To assess the efficacy of our proposal,
we use our conventional rule-based model as a baseline model.
The model queries a store with regular expression-based searching algorithms on specific fields, such as \textit{telephone number} or \textit{address}.
The model may not be able to return a single candidate, for example, franchise stores tend to have the same values in some fields, e.g., \textit{telephone number}, so matching tends to be more difficult.
If multiple entries are returned, predefined rules are applied for re-ranking. For example, the most visited store is returned based on the history.
These heuristic rules are hard to cope with the various exceptions in real-world environments.

\subsubsection{Module Combinations}

As explained in Section 3.1,
we build a grounding model by combining several replaceable modules as explained below.

\paragraph{Backbones.}
To embed the multilingual data,
we test two BERT-based models; Multilingual BERT (MBERT)~\citep{devlin-etal-2019-bert, pires-etal-2019-multilingual} and XLM-Roberta (XLMR)~\citep{conneau-etal-2020-unsupervised}.

\paragraph{Similarity functions.}
To score the similarities between the data (i.e., JSON objects), two similarity functions are tested; inner-product similarity (IPS) $\langle\bs y_{i}, \bs y_{\hat{j}}\rangle$ and negative squared distance (NSD) $-||\bs y_{i} - \bs y_{\hat{j}}||^2_2$. %

\paragraph{Separator token for field concatenation.}
In our pipeline, each JSON object is converted into a sequence of tokens.
During the conversion, we simply concatenate all values in the valid fields separated by a single separator token between the values (Single)~\citep{devlin-etal-2019-bert,karpukhin-etal-2020-dense}.
To emphasize the distinctions between values from different fields,
we also tet multiple field-wise separator tokens (Multi). See Figure~\ref{fig:embedding_process} and Section 3.1 for more details.

\begin{table}[t!]
\centering
\resizebox{\linewidth}{!}{
    \begin{tabular}{ c c c c c }
    \toprule
    \textbf{Backbone} & \textbf{Sim.} & \textbf{Sep.} & \textbf{Mask.} &  \textbf{Acc.} \\ 
    \midrule
    MBERT & IPS & Single & None   & 86.61 \\
    MBERT & IPS & Single & Single & 87.45 \\
    MBERT & IPS & Single & Multi  & 85.38 \\
    MBERT & IPS & Multi  & None   & 88.91 \\
    MBERT & IPS & Multi  & Single & 88.01 \\
    MBERT & IPS & Multi  & Multi  & 88.15 \\
    MBERT & NSD & Single & None   & 85.61 \\
    MBERT & NSD & Single & Single & 87.09 \\
    MBERT & NSD & Single & Multi  & 90.26 \\
    MBERT & NSD & Multi  & None   & 90.85 \\
    MBERT & NSD & Multi  & Single & 90.73 \\
    MBERT & NSD & Multi  & Multi  & \textbf{91.61} \\ %
    XLMR  & IPS & Single & None   & 87.36 \\
    XLMR  & IPS & Single & Single & 89.36 \\
    XLMR  & IPS & Single & Multi  & 90.36 \\
    XLMR  & IPS & Multi  & None   & 90.14 \\
    XLMR  & IPS & Multi  & Single & 89.36 \\
    XLMR  & IPS & Multi  & Multi  & 89.58 \\
    XLMR  & NSD & Single & None   & 89.11 \\
    XLMR  & NSD & Single & Single & 91.05 \\
    XLMR  & NSD & Single & Multi  & 90.94 \\
    XLMR  & NSD & Multi  & None   & 90.55 \\
    XLMR  & NSD & Multi  & Single & 89.87 \\
    XLMR  & NSD & Multi  & Multi  & 89.81 \\
    \bottomrule
    \end{tabular}
}
\caption{\textbf{Performances of all module combinations.} We run each module combination three times and average accuracies are reported in the table. There are considerable performance gaps among the module combinations (2×2×2×3=24 in total).}
\label{tab:module_compare}
\end{table}

\paragraph{Masking token for missing values.}
In \citet{devlin-etal-2019-bert}, a special mask token is used to train the model to capture the associations among input tokens. During the training, the input tokens are randomly replaced by the mask token and the model tries to recover the masked values from its neighbors.
Inspired by this,
we use the mask token to mitigate the negative ramifications of missing values in the data.
When the value is missing, we use either a single masking token (Single) or multiple field-wise masking tokens (Multi) instead of leaving it as blank (None).
See Figure~\ref{fig:embedding_process} and Section 3.1 for more details.

\begin{figure}[t]
    \centering
    \includegraphics[width=\linewidth]{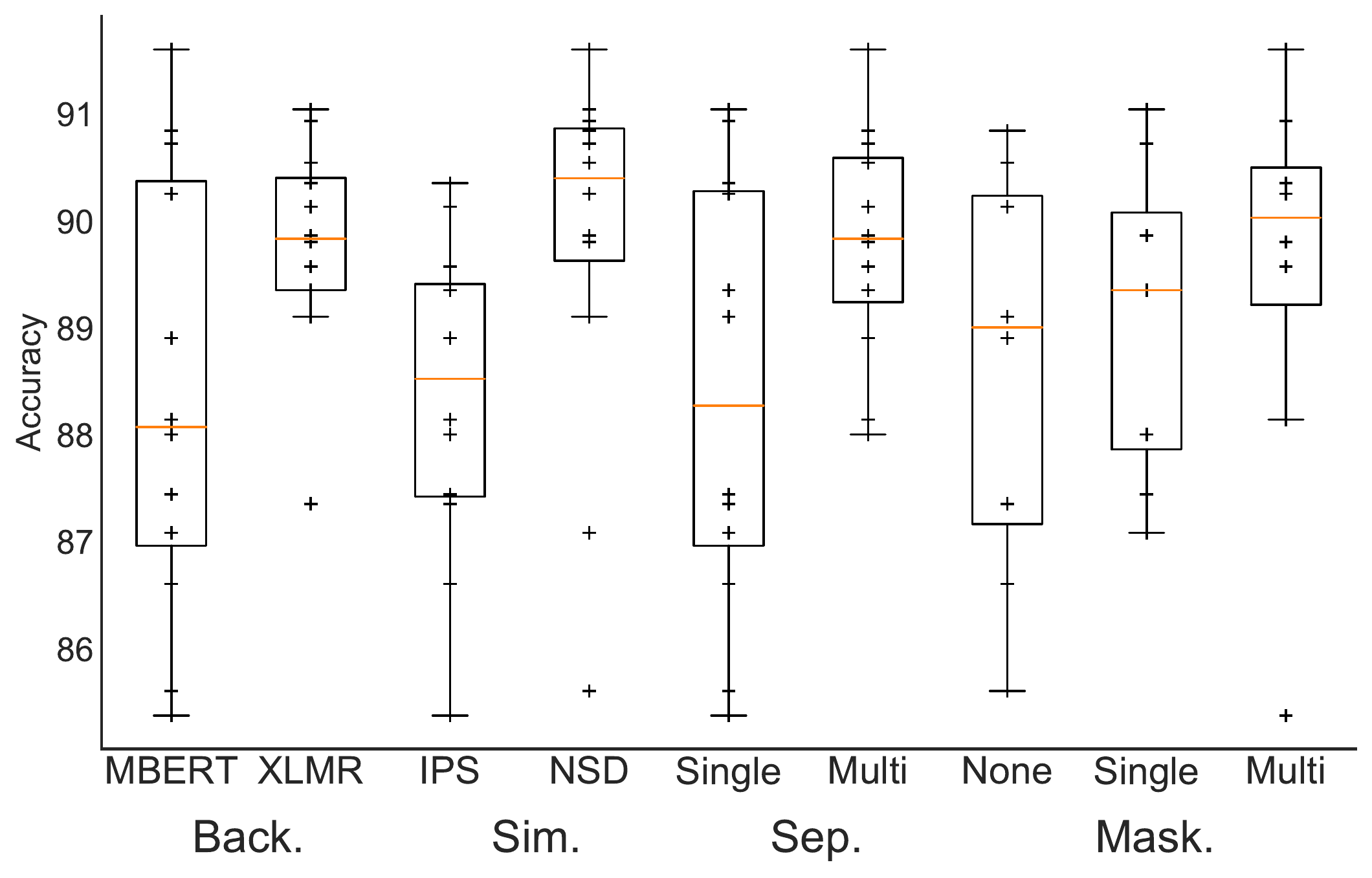}
    \caption{\textbf{Module-wise comparative analysis.} We run each setting three times and average accuracies are reported.}
    \label{fig:module_compare}
\end{figure}

\subsection{Results and Analysis}
We conduct extensive experiments to assess the effectiveness of each module and to find the optimal combination. %

\begin{table}[t]
\centering
\begin{tabular}{l c}
  \toprule
    \textbf{Valid Fields} & \textbf{Acc.} \\
  \midrule
    Store Name        & 33.66 \\
    + Address         & 81.40 \\
    + Telephone Number    & 91.02 \\
    + Business Number & \textbf{91.61} \\
  \bottomrule
\end{tabular}
\caption{\textbf{Analysis on the number of valid fields.} The matching accuracy becomes higher as the number of valid fields grows.} 
\label{tab:valid_field_num}
\end{table}

\paragraph{Experiment 1 (Module Combinations).}
We evaluate all module combinations (2×2×2×3= 24 in total). The results are shown in Table~\ref{tab:module_compare}.
In this experiment, we split the 1M receipt images into test (10K) and train (the rest).
We use the database entries (390K) that are associated with the 1M receipts.
The models are trained to learn the associations between trainset receipts and database entries.
And the trained models are used to predict unobserved associations from the test set receipts.
The batch size is set to 32, the number of steps is 40K, and the learning rate is 2e-5.
As can be seen in Table~\ref{tab:module_compare},
there are considerable performance gaps among the combinations ranging from 85.6 to 91.6.
The best module combination is MBERT-NSD-Multi-Multi.
To show the effect of each module more comprehensively, we plot the module-wise performances in Figure~\ref{fig:module_compare}.
MBERT has a large performance variance, and the combination of MBERT and NSD seems to be effective.
The result also shows that using multiple special tokens (Multi) is effective for mitigating the issues of multiple field concatenation as well as missing values in the data.

\paragraph{Experiment 2 (Valid Field Selection).}
We also investigate the effects of using multiple fields
by changing the number of valid fields during the training.
The results in Table~\ref{tab:valid_field_num} show the matching accuracy becomes higher as the number of valid fields grows.
This highlights the importance of using multiple fields to differentiate similar entries in the database,
leading to a performance improvement in grounding.

\begin{table}[t]
\centering
    \begin{tabular}{ l c  }
    \toprule
    \textbf{Models} & \textbf{Acc.} \\ 
    \midrule
    Baseline Model & 67.0 \\
    Proposed (MBERT-NSD-Multi-Multi) & \textbf{75.3} \\
    \bottomrule
    \end{tabular}
\caption{\textbf{Matching accuracies of the daily queries from the real users in \emph{POI Match}.} The proposed matching algorithm outperforms the heavily engineered baseline method by 8\% point of top-1 matching accuracy.}
\label{tab:comparison}
\end{table}

\paragraph{Experiment 3 (Results on \emph{POI Match}).}
To find out how effective the proposed system is in our application environment, we evaluate the model on the daily queries from real users in \textit{POI Match}.
We gathered 300 queries and manually annotated the correct database entries.
To see the practical gains in the real application scenario,
we use all database entries (approximately 6 million entries) in this experiment, that is, the model should distinctly distinguish numerous similar entries from the full database.
We trained our model with a batch size of 64 for 80K steps.
Table~\ref{tab:comparison} shows, the proposed matching model significantly outperforms the baseline model that is currently deployed by 8\% point of top-1 matching accuracy.
This performance gap corresponds to approximately 2 million user queries per month.

\section{Concluding Remarks and Future Work}
In this paper, we propose a new type of grounding system for querying large-scale document-oriented databases with semi-structured natural language.
The proposed system utilizes embedding-based retrieval to alleviate several practical concerns in the semi-structured query grounding problem.
We examined the proposed system on our application \emph{POI Match} which aims to find the corresponding POI entry for a user input receipt image.
Despite various OCR, Parsing, and DB noises, the proposed method successfully matches the receipt image with the corresponding DB entry. In our experiments, the proposed model significantly outperforms the heavily engineered baseline model that has been used in our product for the last few years while requiring much less development and maintenance cost.

As future work, testing the proposed grounding pipeline on different applications or domains would help to get a general understanding of each module.
Investigating the connections between some recent theoretical analyses on Siamese neural-network-based retrieval and our empirical results would also be interesting.
We believe our findings in this work can easily be extended to other real-world applications that are dependent on a similar problem.

\section*{Acknowledgements}
The authors would like to thank anonymous reviewers for their insightful comments.

\bibliography{anthology,custom}

\begin{thebibliography}{33}
\providecommand{\natexlab}[1]{#1}

\bibitem[{Arora and Aggarwal(2013)}]{arora2013modeling}
Arora, R.; and Aggarwal, R.~R. 2013.
\newblock Modeling and querying data in mongodb.
\newblock \emph{International Journal of Scientific and Engineering Research},
  4(7): 141--144.

\bibitem[{Baek et~al.(2019{\natexlab{a}})Baek, Kim, Lee, Park, Han, Yun, Oh,
  and Lee}]{baek2019STRcomparisons}
Baek, J.; Kim, G.; Lee, J.; Park, S.; Han, D.; Yun, S.; Oh, S.~J.; and Lee, H.
  2019{\natexlab{a}}.
\newblock What Is Wrong With Scene Text Recognition Model Comparisons? Dataset
  and Model Analysis.
\newblock In \emph{International Conference on Computer Vision (ICCV)}.

\bibitem[{Baek et~al.(2019{\natexlab{b}})Baek, Lee, Han, Yun, and
  Lee}]{baek2019craft}
Baek, Y.; Lee, B.; Han, D.; Yun, S.; and Lee, H. 2019{\natexlab{b}}.
\newblock Character Region Awareness for Text Detection.
\newblock In \emph{Proceedings of the IEEE Conference on Computer Vision and
  Pattern Recognition}, 9365--9374.

\bibitem[{Bhardwaj(2016)}]{mongodbcouchdb}
Bhardwaj, N. 2016.
\newblock Comparative Study of CouchDB and MongoDB – NoSQL Document Oriented
  Databases.
\newblock \emph{International Journal of Computer Applications}, 136: 24--26.

\bibitem[{Bojanowski et~al.(2017)Bojanowski, Grave, Joulin, and
  Mikolov}]{bojanowski-etal-2017-enriching}
Bojanowski, P.; Grave, E.; Joulin, A.; and Mikolov, T. 2017.
\newblock Enriching Word Vectors with Subword Information.
\newblock \emph{Transactions of the Association for Computational Linguistics},
  5: 135--146.

\bibitem[{Bromley et~al.(1994)Bromley, Guyon, LeCun, S\"{a}ckinger, and
  Shah}]{siamesenn}
Bromley, J.; Guyon, I.; LeCun, Y.; S\"{a}ckinger, E.; and Shah, R. 1994.
\newblock Signature Verification using a "Siamese" Time Delay Neural Network.
\newblock In Cowan, J.; Tesauro, G.; and Alspector, J., eds., \emph{Advances in
  Neural Information Processing Systems}, volume~6, 737--744. Morgan-Kaufmann.

\bibitem[{Cohan et~al.(2019)Cohan, Beltagy, King, Dalvi, and
  Weld}]{cohan-etal-2019-pretrained}
Cohan, A.; Beltagy, I.; King, D.; Dalvi, B.; and Weld, D. 2019.
\newblock Pretrained Language Models for Sequential Sentence Classification.
\newblock In \emph{Proceedings of the 2019 Conference on Empirical Methods in
  Natural Language Processing and the 9th International Joint Conference on
  Natural Language Processing (EMNLP-IJCNLP)}, 3693--3699. Hong Kong, China:
  Association for Computational Linguistics.

\bibitem[{Conneau et~al.(2020)Conneau, Khandelwal, Goyal, Chaudhary, Wenzek,
  Guzm{\'a}n, Grave, Ott, Zettlemoyer, and
  Stoyanov}]{conneau-etal-2020-unsupervised}
Conneau, A.; Khandelwal, K.; Goyal, N.; Chaudhary, V.; Wenzek, G.; Guzm{\'a}n,
  F.; Grave, E.; Ott, M.; Zettlemoyer, L.; and Stoyanov, V. 2020.
\newblock Unsupervised Cross-lingual Representation Learning at Scale.
\newblock In \emph{Proceedings of the 58th Annual Meeting of the Association
  for Computational Linguistics}, 8440--8451. Online: Association for
  Computational Linguistics.

\bibitem[{Devlin et~al.(2019)Devlin, Chang, Lee, and
  Toutanova}]{devlin-etal-2019-bert}
Devlin, J.; Chang, M.-W.; Lee, K.; and Toutanova, K. 2019.
\newblock {BERT}: Pre-training of Deep Bidirectional Transformers for Language
  Understanding.
\newblock In \emph{Proceedings of the 2019 Conference of the North {A}merican
  Chapter of the Association for Computational Linguistics: Human Language
  Technologies, Volume 1 (Long and Short Papers)}, 4171--4186. Minneapolis,
  Minnesota: Association for Computational Linguistics.

\bibitem[{Ding, Yu, and Hsieh(2019)}]{pmlr-v89-ding19a}
Ding, Q.; Yu, H.-F.; and Hsieh, C.-J. 2019.
\newblock A Fast Sampling Algorithm for Maximum Inner Product Search.
\newblock In Chaudhuri, K.; and Sugiyama, M., eds., \emph{Proceedings of
  Machine Learning Research}, volume~89 of \emph{Proceedings of Machine
  Learning Research}, 3004--3012. PMLR.

\bibitem[{Gillick et~al.(2019)Gillick, Kulkarni, Lansing, Presta, Baldridge,
  Ie, and Garcia-Olano}]{gillick-etal-2019-learning}
Gillick, D.; Kulkarni, S.; Lansing, L.; Presta, A.; Baldridge, J.; Ie, E.; and
  Garcia-Olano, D. 2019.
\newblock Learning Dense Representations for Entity Retrieval.
\newblock In \emph{Proceedings of the 23rd Conference on Computational Natural
  Language Learning (CoNLL)}, 528--537. Hong Kong, China: Association for
  Computational Linguistics.

\bibitem[{Gormley and Tong(2015)}]{elasticsearch}
Gormley, C.; and Tong, Z. 2015.
\newblock \emph{Elasticsearch: The Definitive Guide}.
\newblock O'Reilly Media, Inc., 1st edition.
\newblock ISBN 1449358543.

\bibitem[{{Gupta} and {Rani}(2016)}]{couchdb}
{Gupta}, S.; and {Rani}, R. 2016.
\newblock A comparative study of elasticsearch and CouchDB document oriented
  databases.
\newblock In \emph{2016 International Conference on Inventive Computation
  Technologies (ICICT)}, volume~1, 1--4.

\bibitem[{Huang et~al.(2020)Huang, Sharma, Sun, Xia, Zhang, Pronin,
  Padmanabhan, Ottaviano, and Yang}]{fairembed}
Huang, J.-T.; Sharma, A.; Sun, S.; Xia, L.; Zhang, D.; Pronin, P.; Padmanabhan,
  J.; Ottaviano, G.; and Yang, L. 2020.
\newblock Embedding-Based Retrieval in Facebook Search.
\newblock In \emph{Proceedings of the 26th ACM SIGKDD International Conference
  on Knowledge Discovery \& Data Mining}, KDD '20, 2553–2561. New York, NY,
  USA: Association for Computing Machinery.
\newblock ISBN 9781450379984.

\bibitem[{Hwang et~al.(2019)Hwang, Kim, Seo, Yim, Park, Park, Lee, Lee, and
  Lee}]{hwang2019postocr}
Hwang, W.; Kim, S.; Seo, M.; Yim, J.; Park, S.; Park, S.; Lee, J.; Lee, B.; and
  Lee, H. 2019.
\newblock Post-{\{}OCR{\}} parsing: building simple and robust parser via
  {\{}BIO{\}} tagging.
\newblock In \emph{Workshop on Document Intelligence at NeurIPS 2019}.

\bibitem[{Hwang et~al.(2021{\natexlab{a}})Hwang, Lee, Yim, Kim, and
  Seo}]{hwang-etal-2021-cost}
Hwang, W.; Lee, H.; Yim, J.; Kim, G.; and Seo, M. 2021{\natexlab{a}}.
\newblock Cost-effective End-to-end Information Extraction for Semi-structured
  Document Images.
\newblock In \emph{Proceedings of the 2021 Conference on Empirical Methods in
  Natural Language Processing}, 3375--3383. Online and Punta Cana, Dominican
  Republic: Association for Computational Linguistics.

\bibitem[{Hwang et~al.(2021{\natexlab{b}})Hwang, Yim, Park, Yang, and
  Seo}]{hwang-etal-2021-spatial}
Hwang, W.; Yim, J.; Park, S.; Yang, S.; and Seo, M. 2021{\natexlab{b}}.
\newblock Spatial Dependency Parsing for Semi-Structured Document Information
  Extraction.
\newblock In \emph{Findings of the Association for Computational Linguistics:
  ACL-IJCNLP 2021}, 330--343. Online: Association for Computational
  Linguistics.

\bibitem[{{Johnson}, {Douze}, and {Jégou}(2019)}]{faiss}
{Johnson}, J.; {Douze}, M.; and {Jégou}, H. 2019.
\newblock Billion-scale similarity search with GPUs.
\newblock \emph{IEEE Transactions on Big Data}, 1--1.

\bibitem[{Karpukhin et~al.(2020)Karpukhin, Oguz, Min, Lewis, Wu, Edunov, Chen,
  and Yih}]{karpukhin-etal-2020-dense}
Karpukhin, V.; Oguz, B.; Min, S.; Lewis, P.; Wu, L.; Edunov, S.; Chen, D.; and
  Yih, W.-t. 2020.
\newblock Dense Passage Retrieval for Open-Domain Question Answering.
\newblock In \emph{Proceedings of the 2020 Conference on Empirical Methods in
  Natural Language Processing (EMNLP)}, 6769--6781. Online: Association for
  Computational Linguistics.

\bibitem[{Kim et~al.(2019)Kim, Okuno, Fukui, and Shimodaira}]{wips}
Kim, G.; Okuno, A.; Fukui, K.; and Shimodaira, H. 2019.
\newblock Representation Learning with Weighted Inner Product for Universal
  Approximation of General Similarities.
\newblock In \emph{Proceedings of the Twenty-Eighth International Joint
  Conference on Artificial Intelligence, {IJCAI-19}}, 5031--5038. International
  Joint Conferences on Artificial Intelligence Organization.

\bibitem[{Kim(2014)}]{kim-2014-convolutional}
Kim, Y. 2014.
\newblock Convolutional Neural Networks for Sentence Classification.
\newblock In \emph{Proceedings of the 2014 Conference on Empirical Methods in
  Natural Language Processing ({EMNLP})}, 1746--1751. Doha, Qatar: Association
  for Computational Linguistics.

\bibitem[{Krishnan, Elayidom, and Santhanakrishnan(2016)}]{mongodb}
Krishnan, H.; Elayidom, M.; and Santhanakrishnan, T. 2016.
\newblock MongoDB – a comparison with NoSQL databases.
\newblock \emph{International Journal of Scientific and Engineering Research},
  7: 1035--1037.

\bibitem[{Li et~al.(2008)Li, Ooi, Feng, Wang, and
  Zhou}]{10.1145/1376616.1376706}
Li, G.; Ooi, B.~C.; Feng, J.; Wang, J.; and Zhou, L. 2008.
\newblock EASE: An Effective 3-in-1 Keyword Search Method for Unstructured,
  Semi-Structured and Structured Data.
\newblock In \emph{Proceedings of the 2008 ACM SIGMOD International Conference
  on Management of Data}, SIGMOD '08, 903–914. New York, NY, USA: Association
  for Computing Machinery.
\newblock ISBN 9781605581026.

\bibitem[{Lockard, Shiralkar, and Dong(2019)}]{lockard-etal-2019-openceres}
Lockard, C.; Shiralkar, P.; and Dong, X.~L. 2019.
\newblock {O}pen{C}eres: {W}hen Open Information Extraction Meets the
  Semi-Structured Web.
\newblock In \emph{Proceedings of the 2019 Conference of the North {A}merican
  Chapter of the Association for Computational Linguistics: Human Language
  Technologies, Volume 1 (Long and Short Papers)}, 3047--3056. Minneapolis,
  Minnesota: Association for Computational Linguistics.

\bibitem[{Mikolov, Yih, and Zweig(2013)}]{mikolov-etal-2013-linguistic}
Mikolov, T.; Yih, W.-t.; and Zweig, G. 2013.
\newblock Linguistic Regularities in Continuous Space Word Representations.
\newblock In \emph{Proceedings of the 2013 Conference of the North {A}merican
  Chapter of the Association for Computational Linguistics: Human Language
  Technologies}, 746--751. Atlanta, Georgia: Association for Computational
  Linguistics.

\bibitem[{Nadig, Braschler, and Stockinger(2020)}]{nadig-etal-2020-database}
Nadig, S.; Braschler, M.; and Stockinger, K. 2020.
\newblock Database Search vs. Information Retrieval: A Novel Method for
  Studying Natural Language Querying of Semi-Structured Data.
\newblock In \emph{Proceedings of the 12th Language Resources and Evaluation
  Conference}, 1772--1779. Marseille, France: European Language Resources
  Association.
\newblock ISBN 979-10-95546-34-4.

\bibitem[{Paszke et~al.(2019)Paszke, Gross, Massa, Lerer, Bradbury, Chanan,
  Killeen, Lin, Gimelshein, Antiga, Desmaison, Kopf, Yang, DeVito, Raison,
  Tejani, Chilamkurthy, Steiner, Fang, Bai, and Chintala}]{pytorch}
Paszke, A.; Gross, S.; Massa, F.; Lerer, A.; Bradbury, J.; Chanan, G.; Killeen,
  T.; Lin, Z.; Gimelshein, N.; Antiga, L.; Desmaison, A.; Kopf, A.; Yang, E.;
  DeVito, Z.; Raison, M.; Tejani, A.; Chilamkurthy, S.; Steiner, B.; Fang, L.;
  Bai, J.; and Chintala, S. 2019.
\newblock PyTorch: An Imperative Style, High-Performance Deep Learning Library.
\newblock In Wallach, H.; Larochelle, H.; Beygelzimer, A.; d\textquotesingle
  Alch\'{e}-Buc, F.; Fox, E.; and Garnett, R., eds., \emph{Advances in Neural
  Information Processing Systems}, volume~32, 8026--8037. Curran Associates,
  Inc.

\bibitem[{Pires, Schlinger, and Garrette(2019)}]{pires-etal-2019-multilingual}
Pires, T.; Schlinger, E.; and Garrette, D. 2019.
\newblock How Multilingual is Multilingual {BERT}?
\newblock In \emph{Proceedings of the 57th Annual Meeting of the Association
  for Computational Linguistics}, 4996--5001. Florence, Italy: Association for
  Computational Linguistics.

\bibitem[{Reimers and Gurevych(2019)}]{reimers-gurevych-2019-sentence}
Reimers, N.; and Gurevych, I. 2019.
\newblock Sentence-{BERT}: Sentence Embeddings using {S}iamese {BERT}-Networks.
\newblock In \emph{Proceedings of the 2019 Conference on Empirical Methods in
  Natural Language Processing and the 9th International Joint Conference on
  Natural Language Processing (EMNLP-IJCNLP)}, 3982--3992. Hong Kong, China:
  Association for Computational Linguistics.

\bibitem[{Rigaud et~al.(2019)Rigaud, Doucet, Coustaty, and Moreux}]{post_ocr}
Rigaud, C.; Doucet, A.; Coustaty, M.; and Moreux, J.-P. 2019.
\newblock ICDAR 2019 Competition on Post-OCR Text Correction.
\newblock In \emph{2019 International Conference on Document Analysis and
  Recognition (ICDAR)}, 1588--1593.

\bibitem[{Robertson and Zaragoza(2009)}]{robertson2009probabilistic}
Robertson, S.; and Zaragoza, H. 2009.
\newblock \emph{The probabilistic relevance framework: BM25 and beyond}.
\newblock Now Publishers Inc.

\bibitem[{Tan et~al.(2019)Tan, Zhou, Xu, and Li}]{tan-etal-2019-efficient}
Tan, S.; Zhou, Z.; Xu, Z.; and Li, P. 2019.
\newblock On Efficient Retrieval of Top Similarity Vectors.
\newblock In \emph{Proceedings of the 2019 Conference on Empirical Methods in
  Natural Language Processing and the 9th International Joint Conference on
  Natural Language Processing (EMNLP-IJCNLP)}, 5236--5246. Hong Kong, China:
  Association for Computational Linguistics.

\bibitem[{Wolf et~al.(2020)Wolf, Debut, Sanh, Chaumond, Delangue, Moi, Cistac,
  Rault, Louf, Funtowicz, Davison, Shleifer, von Platen, Ma, Jernite, Plu, Xu,
  Le~Scao, Gugger, Drame, Lhoest, and Rush}]{wolf-etal-2020-transformers}
Wolf, T.; Debut, L.; Sanh, V.; Chaumond, J.; Delangue, C.; Moi, A.; Cistac, P.;
  Rault, T.; Louf, R.; Funtowicz, M.; Davison, J.; Shleifer, S.; von Platen,
  P.; Ma, C.; Jernite, Y.; Plu, J.; Xu, C.; Le~Scao, T.; Gugger, S.; Drame, M.;
  Lhoest, Q.; and Rush, A. 2020.
\newblock Transformers: State-of-the-Art Natural Language Processing.
\newblock In \emph{Proceedings of the 2020 Conference on Empirical Methods in
  Natural Language Processing: System Demonstrations}, 38--45. Online:
  Association for Computational Linguistics.

\end{thebibliography}

\end{document}